\input{aipcheck}

\documentclass[
    ,final            
  ]
  {aipproc}

\layoutstyle{6x9}

\begin{document}

\title{The Incidence of \\ Magnetic Fields in Massive Stars: \\ An Overview of the MiMeS Survey Component}

\classification{97.10.Ld, 97.20.Ec, 97.30.Eh, 97.30.-b}
\keywords      {stars: magnetic fields, stars: early-type, techniques: polarimetric}

\author{J.H. Grunhut}{
  address={Dept. of Physics, Royal Military College of Canada, Kingston, Ontario, Canada}
  ,altaddress={Dept. of Physics, Queen's University, Kingston, Ontario, Canada}
}

\author{G.A. Wade}{
  address={Dept. of Physics, Royal Military College of Canada, Kingston, Ontario, Canada}
}

\author{the MiMeS Collaboration}{
address={}
}

\begin{abstract}
With only a handful of known magnetic massive stars, there is a troubling deficit in the scope of our knowledge of the influence of magnetic fields on stellar evolution, and almost no empirical basis for understanding how fields modify mass loss and rotation in massive stars. Most remarkably, there is still no solid consensus regarding the origin physics of these fields - whether they are fossil remnants, or produced by contemporaneous dynamos, or some combination of these mechanisms. This article will present an overview of the Survey Component of the MiMeS Large Programs, the primary goal of which is to search for Zeeman signatures in the circular polarimetry of massive stars (stars with spectral types B3 and hotter) that were previously unknown to host any magnetic field. To date, the MiMeS collaboration has collected more than 550 high-resolution spectropolarimetric observations with ESPaDOnS and Narval of nearly 170 different stars, from which we have discovered 14 new magnetic stars.

\end{abstract}

\maketitle


\section{Introduction}
Massive stars are those stars with initial masses above about 8 times that of the sun, eventually leading to catastrophic explosions in the form of supernovae. These represent the most massive and luminous stellar component of the Universe, and are the crucibles in which the lion's share of the chemical elements are forged. These rapidly-evolving stars drive the chemistry, structure and evolution of galaxies, dominating the ecology of the Universe - not only as supernovae, but also during their entire lifetimes - with far-reaching consequences. 

Although the existence of magnetic fields in massive stars is no longer in question, our knowledge of the basic statistical properties of massive star magnetic fields is seriously incomplete. 

Prior to the MiMeS Project only a handful of magnetic early B-type stars were known, the majority of these being the lower-mass helium-strong stars \citep[e.g.][]{boh87}, along with only 3 proposed magnetic O-type stars \citep{don02, don06, bour08}. The majority of these stars host a large-scale field, with a strong dipole component.

Due to the difficulty in measuring magnetic fields in massive stars (e.g. due to the broad and relatively few optical spectral lines) the relatively weak magnetic fields in these stars have remained undetected by previous generations of instrumentation. Many new magnetic massive stars have been reported in the recent literature based on low-resolution spectropolarimetric observations obtained primarily with the VLT's FORS instruments. A significant number of those claims have not been confirmed by independent observations \citep[e.g.][]{silv09, shultz11}. In light of these issues it is challenging to interpret the relatively few existing surveys of magnetism in massive stars. This represents a fundamental gap in our understanding of stellar magnetism and the physics of hot stars.

\section{The MiMeS Project}
The Magnetism in Massive Stars (MiMeS) Project represents a comprehensive, multidisciplinary strategy by an international team of recognized researchers to address the ``big questions" related to the complex and puzzling magnetism of massive stars. In 2008, the MiMeS Project was awarded ``Large Program" (LP) status by both Canada and France (PI G. Wade) at the Canada-France-Hawaii Telescope (CFHT), where the Project was allocated 640 hours of dedicated time with the ESPaDOnS spectropolarimeter from late 2008 through 2012. Since then the MiMeS consortium was awarded additional LP status with the Narval spectropolarimeter at the Bernard Lyot Telescope in France (a total of $\sim$500 hours, PI C. Neiner) and the HARPS polarimeter at ESO's 3.6\,m telescope at La Silla, Chile (a total of $\sim$300 hours, PI E. Alecian). In addition to these LPs, the MiMeS Project is supported by numerous PI programs from such observatories as the Anglo-Australia Telescope, Chandra, Dominion Astrophysical Observatory, HST, MOST, SMARTS, and the Very Large Telescope (VLT).

The structure of the ESPaDOnS MiMeS LP includes 255 hours for a $\sim$40-target ``Targeted Component" (TC) which obtains time-resolved high-resolution spectropolarimetery of known magnetic massive stars in unpolarized light (Stokes $I$), circularly polarized light (Stokes $V$), and for some targets also linear polarized light (Stokes $Q$ and $U$), with the goal to constrain detailed models of the surface magnetic field structure and related physics of these stars. In addition, 385 hours is dedicated to the $\sim250$-target ``Survey Component" (SC), with the aim to search for magnetic fields in stars with no prior detection of a magnetic field, to provide critical missing information about field incidence and statistical field properties from a much larger sample of massive stars. The majority of the SC targets were chosen such that they have $V<9$ and a projected rotational velocity $v\sin i<300$\,km\,s$^{-1}$ to ensure sensitivity to relatively weak large-scale fields.

The ESPaDOnS MiMeS SC primarily focuses on Galactic field stars consisting of ``normal" OB stars, emission line B-type and classical Be stars, pre-main sequence Herbig Be stars, and Wolf-Rayet stars. 

The MiMeS Project has four main science objectives:
\begin{itemize}
\item To identify and model the physical processes responsible for the generation of magnetic fields in massive stars;
\item To observe and model the detailed interaction between the magnetic fields and stellar winds of massive stars;
\item To investigate the role of the magnetic field in modifying the rotational properties of massive stars;
\item To investigate the impact of magnetic fields on massive star evolution, and the connection between magnetic fields of non-degenerate massive stars and their descendants, the neutron stars and magnetars.
\end{itemize}

\section{Observations}
The primary instrument used for the SC is the bench-mounted, high throughput, fibre-fed, high-resolution ($R\sim65000$) ESPaDOnS spectropolarimeter. The spectrograph is both thermally and vibrationally isolated to reduce systematic effects, which could result in spurious magnetic signatures. The spectrograph is a cross-dispersed echelle spectrograph capable of obtaining near complete coverage of the optical spectrum (370 to 1050\,nm). The polarimeter is composed of one quarter- and two half-wave Fresnel rhombs with a Wollaston prism to provide achromatic polarization in all four Stokes parameters.  Our exposure times are chosen to statistically provide signal-to-noise ratios (S/N) high enough to detect dipole magnetic fields with a range of surface polar field intensities from 100\,G to 2\,kG depending on the stellar and physical properties.

To diagnose the magnetic field, the SC relies on the magnetic splitting of photospheric spectral lines (the Zeeman effect). Since the magnetic splitting observed in unpolarized light is difficult to detect in hot stars due to their intrinsically broad spectral lines (from thermal, turbulent or rotational broadening), we rely on the Zeeman signatures that are produced in circularly polarized light (Stokes $V$), which can still be observed even in heavily broadened lines. The line splitting observed in the unpolarized light conveys information about the total modulus of the magnetic field strength, but the Zeeman signature in circular polarization only relays information about the line-of-sight (LOS) component of the magnetic field. The relative strength of the Zeeman signature (compared to the continuum level) only depends on the magnetic field strength, the total broadening, and the atomic properties of the spectral line (line depth and magnetic sensitivity). Our ability to resolve these signatures across individual line profiles in metallic and helium lines (as opposed to only hydrogen lines) is only possible because of the high spectral resolution of ESPaDonS, Narval and the HARPs polarimeter. This allows us to unambiguously detect the presence of magnetic signatures, but requires sufficiently high S/Ns in order to distinguish the signatures relative to the noise level. Furthermore, the ability to resolve individual line profiles allows us to (for most magnetic configurations) still detect a Zeeman signature even when the net LOS component of the magnetic field (the longitudinal field $B_\ell$) is consistent with zero. This is not possible with low-resolution instruments that cannot typically resolve individual line profiles, and therefore only rely on the $B_\ell$ measurement for detection purposes.

\begin{figure}
\centering
\includegraphics[width=6in]{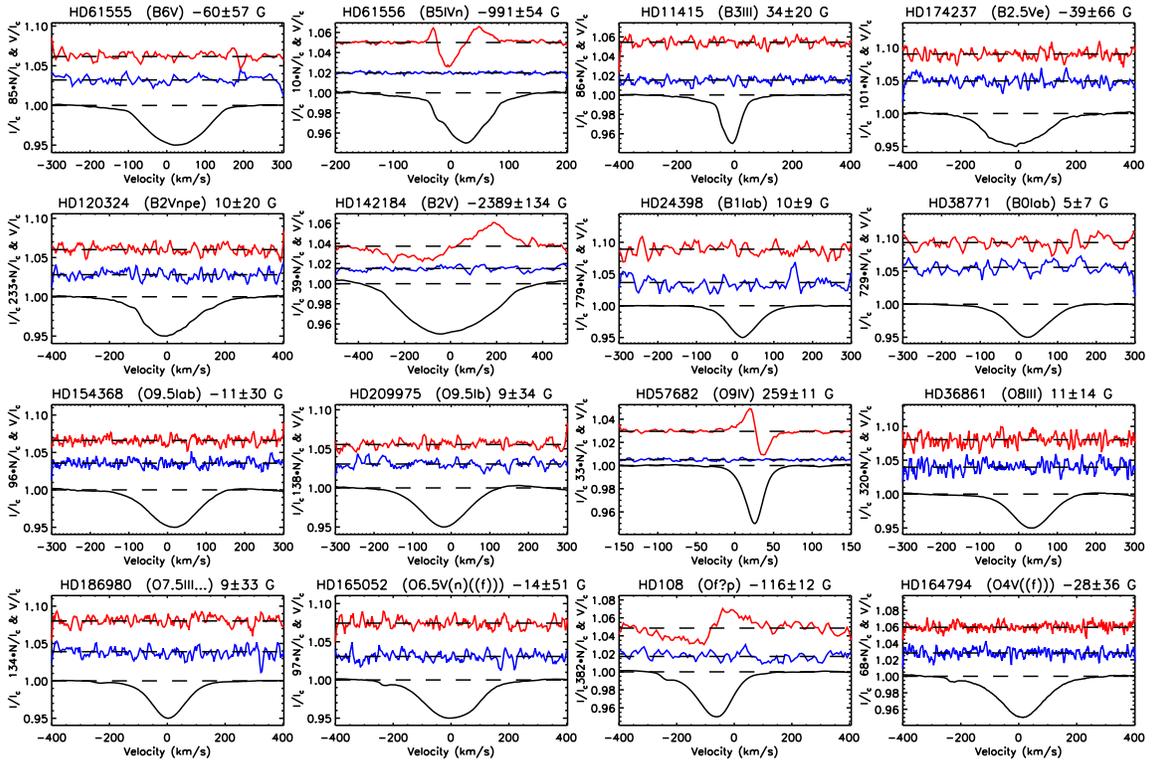}
\caption{Mean LSD circularly polarized Stokes $V$ (top curve, red), diagnostic null (middle curve, blue) and unpolarized Stokes $I$ (bottom curve, black) profiles for a small sample of the Survey Component observations. The spectral lines have all been scaled such that the unpolarized absorption lines have a depth that is 5\% of the continuum and the null and Stokes $V$ profiles have been shifted and amplified by the indicated factor for display purposes. A magnetic field is detected if there is excess polarization signal detected within the line profile of the Stokes $V$ spectrum. Clear magnetic Zeeman signatures are visible in HD\,61556, HD\,142184, HD\,57682, and HD\,108 while no Zeeman signature is present in the other stars. Also included are the spectral type and computed longitudinal field for each LSD profile.}
\label{lsd_examp}
\end{figure}

Since the detected magnetic fields in most hot, massive stars are relatively weak, we require S/Ns on the order of $\sim$10\,000 to detect Zeeman signatures in individual optical spectral lines of typical O- or B-type stars. In order to increase the S/N and enhance our sensitivity to weak Zeeman signatures, we employ the Least-Squares Deconvolution (LSD) multi-line technique of \citet{don97} to produce a single mean profile from the unpolarized light, circularly polarized light and also from the diagnostic spectrum (this null spectrum diagnoses the presence of spurious signatures resulting from instrumental or other systematic effects). The S/N increase is dependent on the number of spectral lines used in the analysis. For comparison with other studies, we also compute the first order moment of Stokes $V$, which was shown to be proportional to $B_\ell$ \citep{rees79}. Example LSD profiles are shown in Fig.~\ref{lsd_examp} for several different SC stars with different spectral types.

\section{Quality Control}
One of the major undertakings of the MiMeS Project is the detailed analysis of previously known magnetic massive stars as part of the TC. Datasets obtained for these stars allow us to constantly assess the reliability of our measurements, which is key to ensuring the results of the SC are trustworthy.

\begin{figure}
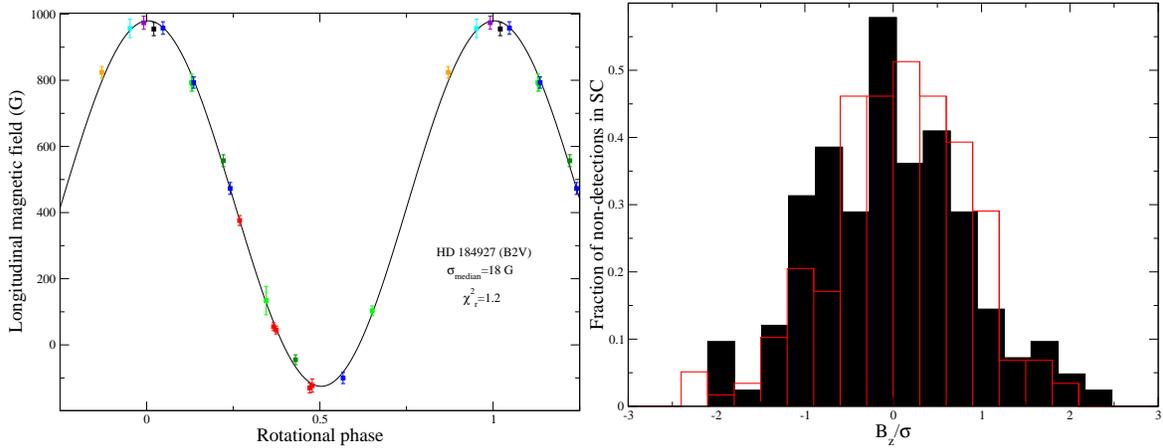

\centering
\includegraphics[width=3in]{jason_hd184927.eps}
\includegraphics[width=3in]{z_hist.eps}
\caption{{\bf Left panel:} Longitudinal magnetic field measurements ($B_\ell$) phased to the rotational period ($\sim9.5$\,d) for HD\,184927, one of the MiMeS TC targets. The colors encode information about the epoch at which the observations were obtained, over the 2 years of study. Note the excellent agreement between observations at different epochs. {\bf Right panel:} Distribution of longitudinal field significances ($B_\ell/\sigma$; solid black) highlighting the expected Gaussian distribution about zero for all the SC observations that have no detectable Zeeman signatures. Also shown is the same distribution computed from the diagnostic null profiles that shows the null distribution is consistent with the Stokes $V$ distribution, which is expected from pure noise.}
\label{qual_cont}
\end{figure}

One key aspect is the reproducibility of our measurements and associated uncertainties (e.g. the Zeeman signatures and corresponding $B_\ell$ measurements). In Fig.~\ref{qual_cont} we display the $B_\ell$ curve for one of the MiMeS TC targets, HD\,184927, which illustrates the repeatability of our measurements over many different epochs, for a star with a rotational period of $\sim$9.5 days, from observations spanning 2 years.

Another quality check that we perform is to test the precision of the instrument. We analyze the distribution of detection significances of the longitudinal magnetic field measurements ($B_\ell/\sigma$), from the observed SC sample with no detectable Zeeman signatures, to ensure that the distribution is symmetric about zero. We also compare this distribution to the same measurements from diagnostic null spectra to confirm they are statistically identical. As shown in the right panel of Fig.~\ref{qual_cont}, we find that the $B_\ell$ measurements are consistent with their error bars and scattered about zero, that we derive a statistically identical distribution from the diagnostic null profiles, and that 75\% and 95\% of our sample have an absolute detection significance of less than 1 and 2, respectively.

Lastly, in the few SC stars that we discover to be magnetic, the target is continuously re-observed to confirm the presence of a field and to verify that the field behaves in a rational manner. Typically, this means that the field is repeatedly detected and that it can be shown to vary periodically \citep[e.g. ][]{grun09,petit11}.

\section{Preliminary Results}
As of July 11th, 2011, the MiMeS SC has observed approximately 166 stars with ESPaDOnS and Narval. This corresponds to 2176 exposures (for certain targets we combine multiple sequences of observations to improve the S/N without saturating the CCD), resulting in 544 polarized spectra (4 exposures are required for a single polarimetric observation). This consumed roughly 228 hours of telescope time. The data are of extremely high precision, with a mean peak S/N per 1.8\,km\,s$^{-1}$ pixel in the unpolarized spectra of $\sim$1300. After performing a preliminary LSD analysis of these spectra, we measure an average S/N of $\sim$14\,000 in the mean LSD Stokes $V$ profiles.

\begin{figure}
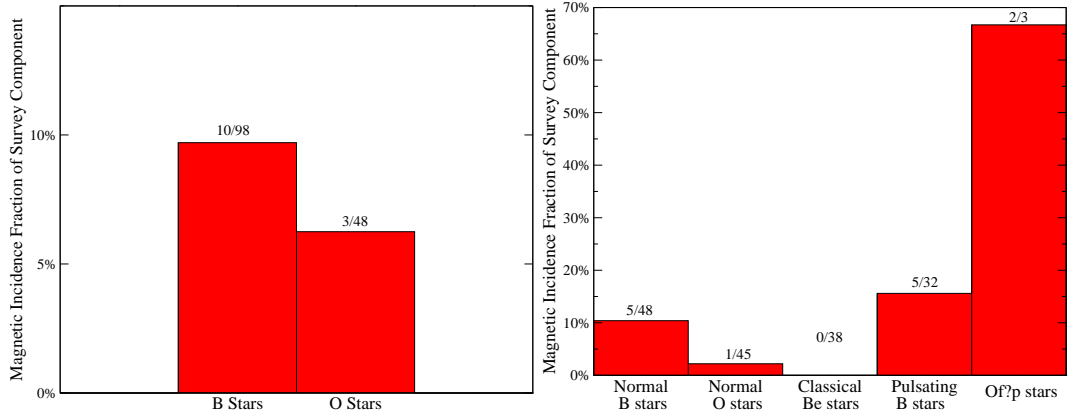

\centering
\includegraphics[width=2.75in]{cat_type_det_hist2.eps}
\includegraphics[width=2.75in]{cat_type_det_hist.eps}
\caption{Incidence fraction of magnetic stars relative to the total indicated sample of all stars observed as part of the MiMeS survey and related PI programs. Our statistics indicate a $\sim$8\% incidence rate of magnetism amongst all massive stars in our sample.}
\label{stat_res}
\end{figure}

From our preliminary statistics we find that approximately 8\% of all O- and B-type stars observed as part of the SC and related PI programs host detectable magnetic fields. If we look at the incidence rate of magnetic fields amongst just O- or B-type stars (left panel of Fig.~\ref{stat_res}) we find that $\sim$10\% of B-type stars host magnetic fields, while only 6\% of O-type stars are found to be magnetic. 

If we further classify the observed stars based on their spectroscopic or other fundamental properties (right panel of Fig.~\ref{stat_res}) we find that the peculiar class of massive, emission line Of?p stars stands out, with roughly 66\% (2/3) of these stars observed to be magnetic. These are small number statistics but if we include another Of?p star, which was previously found to be magnetic  \citep[HD\,191612,][]{don06}, and is being observed as part of the TC, we find that 3/4 of these stars that were observed with ESPaDOnS are found to be magnetic. Given this high rate of detection, it may well be that a field was not detected in the other observed star as this star was too faint to achieve a high enough S/N to detect a similarly weak field as found in the other Of?p stars.

We also point out the class of pulsating B-type stars that consists of the Slowly Pulsating B-type (SPB) stars and the $\beta$ Cep pulsators. These stars are of particular interest since it has been claimed that upwards of 40\% of these stars have been found to be magnetic from low-resolution observations \citep[e.g.][]{hub06, hub09}. Our measurements are in conflict with these results as we only find detectable magnetic fields in about 16\% of the observed stars, which is similar to the incidence fraction ($\sim$10\%) that we obtain in normal B-type stars. For a few of these stars we have obtained multiple observations and have carried out a more detailed analysis \citep{shultz11}.

Lastly, another interesting preliminary result is the lack of detected fields in any classical Be star. We follow the definition of classical Be stars as B-type stars close to the main sequence that exhibit line emission over the photospheric spectrum, resulting from circumstellar gas that is confined to an equatorial decretion disk that is rotating in a Keplerian orbit \citep{port03}. This is in contrast to the common Be star classification, which is typically defined as a B-type star that has shown emission during at least one point in its life. This distinction allows us to investigate the potential role of magnetic fields in the formation of such circumstellar disks.

The lack of detections in classical Be stars cannot be a result of systematically achieving a lower sensitivity in these stars compared to the normal B-type stars. We do find, on average, that the classical Be stars have higher rotational velocities compared to the normal B-type stars, but despite this fact, we still sample a population of classical Be stars for which we should have detected at least a few magnetic stars, assuming that classical Be stars have a magnetic incidence fraction  ($\sim$10\%) and magnetic field properties similar to the normal B-type stars. We also confirm that the lack of detections is not a result of emission contributions since we detect magnetic fields in other emission line O- and B-type stars (that do not fall into the classical Oe or Be star classification, e.g. Of?p stars), which indicates we are capable of observing Zeeman signatures in stars with emission.

\section{Summary}
In summary, the MiMeS collaboration has discovered a total of 14 new magnetic massive stars as part of the MiMeS LP. Of these 14 stars, 3 are O-type stars, which doubles the number of magnetic O-type stars known prior to the start of the MiMeS Project. Our preliminary statistics suggest that $\sim$8\% of all hot, massive stars host strong, large-scale magnetic fields, which qualitatively agrees with the 5-10\% incidence rate that is found for intermediate mass stars \citep{power07} suggesting a similar mechanism for the magnetic field generation. We do not detect large-scale magnetic fields in any classical Be star, likely indicating that strong, large-scale fields do not take part in the generation of the equatorial decretion disks. Lastly, we tentatively suggest that we have identified a class of magnetic O-type stars, the Of?p stars, with fields being detected in all stars of this class observed with reasonable S/Ns.


  



\bibliographystyle{aipproc}   

\clearpage

\noindent {\bf M. Magalhaes:} I noticed your sample does not include any B[e] supergiants. Is there any reason for that?\\
{\bf JHG:} Not really. Our interest in focusing on classical Be stars was to try and understand the role that magnetic fields could play in the formation and stabilization of a decretion disk. Our sample does include a number of supergiants though.\\
\\
{\bf J. Brown:} When you say ``magnetic massive stars are rare," what roughly is the upper limit on the B-field value? \\
{\bf JHG:}  We have not investigated this in detail yet, but it appears that other than $\zeta$ Ori A \citep{bour08}, all other magnetic massive stars have surface dipole field strengths above 300\,G, consistent with the findings of Auri\`{e}re et al. (2008) for cooler magnetic Ap stars.\\
\\
{\bf J. Cassinelli:} You show results for HR\,5907.  It is a rapidly rotating B star.  You also show that H$\alpha$ is in emission.  That fact is the definition of ``Be star" -- i.e., a B star that has ever shown H$\alpha$ in emission.  Why do you say that it is not a Be star?  You also say that it is not a classical Be star, because it introduces a magnetic field.  It seems that you are thereby excluding the possibility that Be stars are associated with magnetic fields.  You cannot mix theoretical modeling ideas with stellar classification criteria.  In the case of the ``Be" stars, ``H alpha in emission in a near-main sequence star" is the identifying criterion.\\
{\bf JHG:} HR\,5907 is a Be star since it shows emission in hydrogen lines. In fact, many of our magnetic stars show strong and variable emission too. However, HR\,5907's (and the other magnetic emission line stars') H$\alpha$ line profile is morphologically different than the lines of classical Be stars - the emission is not found to be in Keplerian orbit. This is the distinguishing factor and reason that HR\,5907 is not considered a {\it classical} Be star.\\
{\bf Th. Rivinius:} Comment on Cassinelli's question about whether HR\,5907 would be a classical Be star.  While meeting the plain ``main sequence B star with emission" definition, we must acknowledge that such a classification, based on a single morphological property, not necessarily distinguishes separate physical object classes.  Actually, apart from the plain presence of emission, HR\,5907 and HR\,7355 are different from the main bulk of classical Be stars in almost every other property.  It is no more a classical Be star than $\sigma$ Ori E is.  Actually, it has happened frequently that formerly ``classical Be" stars have been awarded their own class once we have learned enough, such as Herbig Be stars or Algols.  See \citet{port03} for a more complete discussion.

\end{document}